\documentclass[12pt]{article}
\usepackage[T1]{fontenc}
\usepackage[latin9]{inputenc}

\usepackage{babel,color}
\usepackage{amsmath, amssymb}

\newtheorem{Theorem}{Theorem}[section]

\newtheorem{Proposition}[Theorem]{Proposition}
\newtheorem{Definition}[Theorem]{Definition}
\newtheorem{rem}[Theorem]{Remark}

\newcommand{\R}{\mathbb{R}}

\newcommand{\N}{\mathbb{N}}

\newcommand{\ii}{\operatorname{ii}} 
\title{On Strict Ranking by Pairwise Comparisons}
\author{Jean-Pierre Magnot}

\begin{document}
	
	\maketitle
	
	\begin{center}
		\it CNRS, LAREMA, SFR MATHSTIC, F-49000 Angers, France, \\ Lepage Research Institut, 17. novembra 1
		081 16 Presov,
		Slovakia\\
		and \\
		Lyc\'ee Jeanne d'Arc, Avenue de Grande Bretagne, F-63000 Clermont-Ferrand
	\end{center}
	
	\begin{center}
		\it jean-pierre.magnot@ac-clermont.fr, magnot@math.cnrs.fr
	\end{center}
	
	\begin{abstract}
		This paper addresses the problem of obtaining a strict ranking (i.e., a ranking without equally ranked items) of \( n \) items based on a pairwise {{}, comparisons matrix}. The basic structures {{}, are described} and a heuristic approach based on a condition, the \(\mathcal{R}\)-condition {{}, is proposed.} The limitations of this ranking procedure {{}, are analyzed}.
	\end{abstract}
	
	\noindent
	\textbf{Keywords:} Ranking, inconsistency, pairwise comparisons.
	
	\noindent
	\textit{MSC (2020):} 90B50, 81T13, 91B06
	
	\section{Introduction}
	The ranking problem in the {{}, pairwise comparions method (PC method)} is a central challenge when prioritizing or selecting alternatives based on multiple criteria. {{},PC method} involves comparing alternatives using pairwise comparisons, but when inconsistencies arise in these comparisons, the process of deriving a definitive ranking becomes challenging. This issue can be caused by human judgment biases, conflicting comparisons, or the inherent complexity of the problem. Resolving this ranking issue is critical to ensuring that the PC method produces meaningful and accurate decision outcomes.
	
	This problem has been widely studied in theoretical literature (e.g., \cite{BR2022,BR2008,CW2004,6,10,F2008,K1993,KMal2017,KO1997,KSS2020,KGDW2015,27,Ma2018-1,Ma2018-3,MMC2023,M2024,S1977}) and applied to various fields, including asset management and finance \cite{FV2012,I2000}, wireless networks \cite{LLA2012}, and more. However, the issue of obtaining a strict ranking without equally ranked alternatives remains an open problem. In this paper, we propose a mathematical approach to address this issue, leveraging the topology of positive real numbers and the concept of finite configurations.
	
	Our contributions include the introduction of the \(\mathcal{R}\)-condition, which ensures that a strict ranking is achievable without {{}, consistency in the PC matrix} (regardless its concrete meaning). {{},They also provide} a study on the robustness of this ranking method under a chosen way to produce a consistent PC matrix from a inconsistent one. {color{red} They finally suggest} a minimization problem that can be applied to any pairwise {{}, comparisons matrix}. {{}, We prove that this last problem procuces only consistent pairwise comparisons matrices that produce a strict ranking.}
	
	The paper is structured as follows:
	\begin{itemize}
	\item In Section \ref{s1}, we introduce preliminary results on pairwise comparison matrices, inconsistency indices, and {{}, procedures making consistent a PC matrix.}
	\item In Section \ref{s2}, we explore the structures in pairwise comparisons that lead to a strict ranking and discuss the limitations of consistencization {{}, procedures making consistent a PC matrix.}
	\item In Section \ref{s3}, we propose a minimization problem to achieve non-equal ranking and outline potential methods for solving it.
	\item {{}, In conclusion, we address} the meaning of the mathematical structures described in this {{}, work, and} we give in the appendix a short description of a mathematical object, called finite configurations, which reminiscently appeared as a shadow behind the investigations that led to the production of this paper. 
	 \end{itemize}
	\section{Preliminaries} \label{s1}
	{{}, Let $n$ be an integer, $n \geq 3.$} A {{},$n-$}pairwise comparison (PC) matrix \( (a_{i,j}) \) is a \( n \times n \) matrix with entries in \( \mathbb{R}_+^* = \{ x \in \mathbb{R} \mid x > 0 \} \), such that for all \( i,j \), we have \( a_{j,i} = a_{i,j}^{-1} \). The set of \( n \times n \) PC matrices is denoted by \( \text{PC}_n \). For the rigor and the fluidity of the notations, we will use extensively the Bourbaki's notation $\N_n = \{1,\cdots,n\}$ for $n \in \N^*.$ 
	
	Inconsistencies in pairwise comparisons can be explained using cycles of three comparisons, called triads, such as \( (x,y,z) \), where the transitivity property is violated, i.e., \( x \cdot z \neq y \), which leads to inconsistency. To measure inconsistency, we typically define inconsistency indicators, which are mappings from \( \text{PC}_n \) to \( \mathbb{R}_+ \).
	
	\begin{rem}
		Inconsistency, in this context, refers to a measure of inconsistency, not the concept itself. A consistent matrix satisfies the relation \( a_{i,j} \cdot a_{j,k} = a_{i,k} \).
	\end{rem}
	
	A consistent {{},$n-$}PC matrix (denoted by \( \text{CPC}_n \)) allows us to assign weights \( (w_i)_{i \in \mathbb{N}_n} \) to the items, such that {{}, \( a_{i,j} = \frac{w_i}{w_j} \).} These weights provide a numerical ranking of the alternatives.
	
	One efficient method for minimizing a functional is the gradient method, which has been successfully applied to inconsistency indicators in \cite{MMC2023}. The gradient method identifies the direction in which the inconsistency decreases most rapidly. By applying the gradient method, we can make the necessary adjustments to the initial PC matrix to minimize inconsistency and achieve a strict ranking. However, as already observed in \cite{MMC2023}, confirming the first observations of e.g. \cite{CW2004} on less refined settings, two {{}, procedures making consistent a PC matrix.} can lead to different rankings. 
	
	\subsection{From Multiplication to Addition: A Key Approach for Making a PC Matrix Consistent}
	The logarithmic map \( \ln \) is a morphism of groups from the multiplicative group \( (\mathbb{R}_+^*,.) \) to the additive group \( (\mathbb{R},+) \). This correspondence allows us to transform PC matrices expressed in multiplication into those expressed in addition. For example, the multiplicative PC matrix
	
	\[
	\left( \begin{array}{ccc}
		1 & e^a & e^b \\
		e^{-a} & 1 & e^c \\
		e^{-b} & e^{-c} & 1
	\end{array} \right)
	\]
	
	can be transformed into the corresponding additive PC matrix
	
	\[
	\left( \begin{array}{ccc}
		0 & a & b \\
		-a & 0 & c \\
		-b & -c & 0
	\end{array} \right).
	\]
	
	The transformation to addition simplifies the process of making a matrix consistent in some already existing {{}, procedures making consistent a PC matrix}, in particular in the method described in \cite{KO1997}, in which the orthogonal projection method applied to additive PC matrices helps achieve consistency.
	
	\subsection{Lie Theory and Loop Quantum Gravity: Insights into Pairwise Comparison Matrices}
	From a Lie theory perspective, \( \mathbb{R}_+^* \) is an Abelian Lie group, and its tangent space at \( 1 \) corresponds to its Lie algebra, which can be identified with \( \mathbb{R} \). The exponential map \( \exp: \mathbb{R} \to \mathbb{R}_+^* \) is a smooth diffeomorphism. In the context of quantum gravity and Yang-Mills theory, the minimization of the distance between holonomies and \( 1 \) is of interest. We apply these concepts to pairwise comparisons by considering the logarithmic transformation. This approach is described in \cite{Ma2018-1,Ma2018-3}.
	
	In the following sections, we continue by detailing the minimization problem that leads to a strict ranking without equal items. This problem is formulated to allow the use of gradient methods for efficient solution computation.
	\section{Strict Ranking, Related Sets of Weights, and Pairwise Comparisons} \label{s2}
	
	A \emph{strict ranking} is defined by a family of weights $(w_i)_{i \in \mathbb{N}_n}$, where these weights define an injective map $\mathbb{N}_n \rightarrow \mathbb{R}+^*$. This is equivalent to the condition $(w_i)_{i \in \mathbb{N}_n} \in O\Gamma_n({{}, \mathbb{R}_+^*})$, as detailed in the appendix. Moreover, let $\mathfrak{S}_n$ denote the group of bijections of the set $\mathbb{N}_n$. The condition of strict ranking is then equivalent to the existence of a permutation $\sigma \in \mathfrak{S}_n$ such that the weights satisfy:
	$$ w_{\sigma(1)} < \cdots < w_{ \sigma(n)}$$
	
	\noindent
	and the corresponding consistent pairwise comparison (PC) matrix with coefficients {{}, $a_{i,j} = w_i / w_j$} must satisfy the ranking condition:
	
	\begin{equation} \label{Req} i = j \quad \Leftrightarrow \quad a_{i,j} = 1. \end{equation}

	\noindent
	This condition is referred to as the \emph{ranking condition} or the $\mathcal{R}$-condition.
	
	\begin{Definition}
		For $n \geq 3$, let $\mathcal{R}PC_n$ denote the set of $n \times n$ pairwise comparison matrices (which are not necessarily consistent) that satisfy the $\mathcal{R}$-condition, and let $\overline{\mathcal{R}}CPC_n$ represent the set of $n \times n$ consistent pairwise comparison matrices that do not satisfy this condition.
	\end{Definition}

	\subsection{Ranking Loci in the $\mathcal{R}$-Condition and Non-Consistent Ranking}
	
	Since $\mathbb{R}_+^* - \{1\}$ has two connected components, the set
	$$ \mathcal{R}PC_n = \left\{ (a_{i,j})_{(i,j)\in \N_n^2} \in PC_n \, | \, i = j  \Leftrightarrow  a_{i,j} =1 \right\}$$
	\noindent
	has $2^{\frac{n(n-1)}{2}}$ connected components, which are characterized by the set of indices
		$$ I((a_{i,j})_{(i,j)\in \N_n^2}) = \left\{ (i,j)\in \N_n^2 \, | \, i < j \hbox{ and } a_{i,j}>1\right\},$$ or equivalently by 
		$$ J((a_{i,j})_{(i,j)\in \N_n^2}) = I - \N_n^2 = \left\{ (i,j)\in \N_n^2 \, | \, i < j \hbox{ and } a_{i,j}<1\right\},$$
	
	It is important to note that a \emph{consistent} pairwise {{}, comparisons matrix} does not uniquely define a family of weights $(w_i)_{i \in \mathbb{N}_n}$. However, a necessary and sufficient condition for obtaining a strict order
$$ w_{\sigma(1)} < \cdots < w_{ \sigma(n)},$$
	\noindent
	with the corresponding {{}, permutation $\sigma \in \mathfrak{S}_n,$} is that $$ \forall (i,j) \in \N_n, i < j \Leftrightarrow {{}, \frac{w_{\sigma(i)}}{w_{\sigma(j)}}} = a_{\sigma(i),\sigma(j)} < 1.$$
	\noindent
	This leads to the following definitions:
	
	\begin{Definition} \label{d:admissible} An admissible locus for strict ranking in $\mathcal{R}PC_n$ is one of its connected components constituted of matrices $(a_{i,j})_{(i,j)\in \N_n^2} \in PC_n$ such that:
		$$\exists \sigma \in \mathfrak{S}_n, \forall (i,j) \in \N_n^2, i<j \Leftrightarrow a_{\sigma(i)\sigma(j)} < 1.$$
	\end{Definition}
	
	In this definition, consistency is not assumed. Even with inconsistent comparisons, it is clear that the item indexed by $\sigma(i)$ is ranked lower than the item indexed by $\sigma(i+1)$ for each $i \in \mathbb{N}_{n-1}$, since $a_{\sigma(i), \sigma(i+1)} < 1$. Furthermore, this ranking property is preserved between the items indexed by $\sigma(i)$ and $\sigma(i+p)$ for $i \in \mathbb{N}_{n-1}$ and $p \in \mathbb{N}_{n-i}$, as we also have $a_{\sigma(i), \sigma(i+p)} < 1$. Hence, strict ranking appears to be unrelated to consistency in this approach, at least in a purely mathematical viewpoint. In other words, \emph{we have derived an order between items from a non-necessarily consistent PC matrix}, as long as this matrix lies within an admissible locus for strict ranking in $\mathcal{R}PC_n$.
	
	{{},
	\begin{Theorem} \label{R-index} If $n \geq 3,$ here exists a bijection between $\mathfrak{S}_n$ and the admissible loci in $\mathcal{R}PC_n$. \end{Theorem} 
	
	\noindent
	\textbf{Proof.} We now analyse deeper Definition \ref{d:admissible}, and in particular the existence of a permutation  $ \sigma \in \mathfrak{S}_n $ such that $\forall (i,j) \in \N_n^2, i<j \Leftrightarrow a_{\sigma(i)\sigma(j)} < 1.$

	We first make an observation. If two admissible matrices $(a_{i,j})_{(i,j) \in \N_n^2}$ and $(b_{i,j})_{(i,j) \in \N_n^2}$ are in the same locus, $a_{i,j} < 1 \Leftrightarrow b_{i,j} < 1.$ Therefore, if $\sigma \in \mathfrak{S}_n$ is such that 
	 $$\forall (i,j) \in \N_n^2, i<j \Leftrightarrow a_{\sigma(i)\sigma(j)} < 1,$$
	 this is equivalent to state that 
	  $$\forall (i,j) \in \N_n^2, i<j \Leftrightarrow b_{\sigma(i)\sigma(j)} < 1.$$
	In other words, Definition \ref{d:admissible} defines a mapping from $\mathfrak{S}_n$ to the set of admissible loci which is surjective.  
	
	Let $A=(a_{ij})_{(i,j)\in \N_n^2}$ be a PC matrix in an admissible locus. Let $\sigma \in \mathfrak{S}_n$ related to $A.$  We claim that this permutation $\sigma$ is unique. Indeed, let us assume that there exists two such permutations $\sigma$ and $\sigma',$ with $\sigma \neq \sigma'.$ Then $\forall i \in \N_{n-1}, a_{\sigma(i) \sigma(i+1)} < 1$ and $a_{\sigma'(i) \sigma'(i+1)} < 1. $
	Building by induction the weights 
	$$ \left\{ \begin{array}{cl} w_\sigma(1) & = 1 \\
	w_{\sigma(i+1)} & = a_{\sigma(i+1) \sigma(i)} w_\sigma(i) \end{array} \right.$$
	and 
	$$ \left\{ \begin{array}{cl} w'_\sigma(1) & = 1 \\
	w'_{\sigma(i+1)} & = a'_{\sigma(i+1) \sigma(i)} w'_\sigma(i) \end{array} \right.$$
	we get
	$$ \forall i < j, \quad w_{\sigma(i)} < w_{\sigma(j)} \hbox{ and } w'_{\sigma'(i)} < w'_{\sigma(j)}$$
	Let $ k \in \N_n$ be the minimal element such that $w_k \neq w'_k.$ This element existst because $\sigma \neq \sigma',$ and $w_1 = w'1 = 1 \Rightarrow k \geq 2.$
	    Then 
	$$  \sigma' \circ \sigma^{-1}(k) \notin \N_{k-1},$$ 
	$$   \sigma \circ \sigma'^{-1}(k) \notin \N_{k-1},$$
	and $$\sigma^{-1}(k) \neq \sigma'^{-1}(k).$$
	Therefore 
	\begin{itemize}
	\item either $\sigma^{-1}(k) < \sigma'^{-1}(k)$ then $ \sigma^\circ \sigma^{-1}(k) = k < \sigma \circ \sigma'(k)$ and $a_{\sigma^{-1}(k)\sigma'^{-1}(k)}<1.$ But now, $w'_{\sigma' \circ \sigma(k)} < w'(k)$ and ${\sigma' \circ \sigma(k)} \notin \N_{k-1},$ which is impossible.  
	\item or $\sigma^{-1}(k) > \sigma'^{-1}(k)$ and the same arguments hold, exchanging $\sigma$ and $\sigma',$ $w$ and $w'$ in the previous item.
	\end{itemize}
	This shows that the existence of two distinct permutations $\sigma$ and $\sigma'$ to a same matrix $A$ in an admissible locus is impossible. 
	
	We conclude that the mapping from $\mathfrak{S}_n$ to the set of admissible loci is injective, and hence bijective.   
	}
 \begin{Definition}
	Let $(a_{i,j})_{(i,j)\in \N_n^2} \in PC_n.$ The characteristic ranking matrix of $(a_{i,j})_{(i,j)\in \N_n^2}$ is the additive, maybe inconsistent, PC matrix $(c_{i,j})_{(i,j)\in \N_n^2}$ where $$ c_{i,j} = \hbox{\rm sign}(\log(a_{i,j})) =  \left\{\begin{array}{ccl} 0 & \hbox{ if } & a_{i,j} = 1 \\
		1 & \hbox{ if } & a_{i,j} > 1 \\
		-1 & \hbox{ if } & 0<a_{i,j} <1 \end{array} \right. . $$
\end{Definition}

		The characteristic ranking matrix naturally characterizes the connected components of $\mathcal{R}PC_n$, which justifies the terminology used.
		
		\begin{Theorem} \label{char-matrix} There exists a bijection between the set of charactristic matrices and the set of loci (not necessarily admissible) in $\mathcal{R}PC_n$. \end{Theorem}
		
		\noindent
		\textbf{Proof.} There is a bijection between $\mathfrak{S}_n$ and all the possible rankings of $n$ items. Given a ranking 
		$$ w_{\sigma(1)} < \cdots < w_{\sigma(n)},$$
		we get  the corresponding characteristic ranking matrix by $$c_{i,j} = {{}, \hbox{\rm sign}(\log(w_i) - \log(w_j)) = \hbox{\rm sign}(w_i - w_j).}$$
		\vskip 12pt
		
		From this result, we deduce the following:
		
		\begin{Theorem} For all $n \geq 3$, there {{}, exists non-admissible loci} in $\mathcal{R}PC_n$. \end{Theorem}
		
		\noindent \textbf{Proof.} The cardinality of $\mathfrak{S}_n$ is $n!$, while the cardinality of loci in $\mathcal{R}PC_n$ is $2^{n(n-1)/2}$. For $n {{}, \geq} 3$, $n! \neq 2^{n(n-1)/2}$, so there is no bijection between $\mathfrak{S}_n$ and loci in $\mathcal{R}PC_n$. Therefore, by Theorem \ref{R-index}, the result follows.
		
		\vskip 12pt
		
		Let us now analyze the case of $3 \times 3$ PC matrices with the $\mathcal{R}$-condition:
		
		\begin{Theorem} Not all loci of $\mathcal{R}PC_3$ are admissible loci. Among the 8 loci, 6 are admissible. The loci of $\mathcal{R}PC_3$ that are not admissible are those matrices where $a_{1,3}$, $a_{2,1}$, and $a_{3,2}$ are simultaneously either in $(1, +\infty)$ or in $(0, 1)$. \end{Theorem}
		
	\noindent
	\textbf{Proof.}
	A PC-matrix in $\mathcal{R}PC_3$has: 
	\begin{itemize}
		\item its diagonal entries equal to 1
		\item three entries less than 1
		\item three entries bigger than 1
	\end{itemize} 
	We analyze characteristic ranking matrices. Let us check all cases:
	\begin{itemize}
		\item For  $\left(\begin{array}{ccc} 0 & -1 & -1 \\
			1 & 0 & -1 \\
			1 & 1 & 0 \end{array}\right),$ we get directly $w_1 < w_2 < w_3.$
		\item For $\left(\begin{array}{ccc} 0 & 1 & 1 \\
			-1 & 0 & 1 \\
			-1 & -1 & 0 \end{array}\right),$ the permutation $\sigma = \left(\begin{array}{cc} 1 & 3  \end{array}\right)$ gives $w_3 < w_2 < w_2.$
		
		\item by the action of $\sigma = \left(\begin{array}{cc} 1 & 2  \end{array}\right)$ 	and $\sigma' = \left(\begin{array}{cc} 2 & 3  \end{array}\right)$ on  $\left(\begin{array}{ccc} 0 & -1 & -1 \\
			1 & 0 & -1 \\
			1 & 1 & 0 \end{array}\right),$ we get respectively the matrices   $\left(\begin{array}{ccc} 0 & 1 & -1 \\
			-1 & 0 & -1 \\
			1 & 1 & 0 \end{array}\right)$ and $\left(\begin{array}{ccc} 0 & -1 & -1 \\
			1 & 0 & 1 \\
			1 & -1 & 0 \end{array}\right),$ which shows that the corresponding loci are admissible.
		\item by the action of the cyclic permutations  $\sigma = \left(\begin{array}{ccc} 1 & 2 & 3  \end{array}\right)$ 	and $\sigma' = \left(\begin{array}{ccc} 1 & 3 & 2 \end{array}\right)$ on $\left(\begin{array}{ccc} 0 & -1 & -1 \\
			1 & 0 & -1 \\
			1 & 1 & 0 \end{array}\right),$ we get respectively the matrices   $\left(\begin{array}{ccc} 0 & -1 & 1 \\
			1 & 0 & 1 \\
			-1 & -1 & 0 \end{array}\right)$ and $\left(\begin{array}{ccc} 0 & 1 & 1 \\
			-1 & 0 & -1 \\
			-1 & 1 & 0 \end{array}\right),$ which shows that the corresponding loci are admissible.
		\item There are two loci not attained by the action of $\mathfrak{S}_3,$ which are represented by $\left(\begin{array}{ccc} 0 & 1 & -1 \\
			-1 & 0 & 1 \\
			1 & -1 & 0 \end{array}\right)$ and $\left(\begin{array}{ccc} 0 & -1 & 1 \\
			1 & 0 & -1 \\
			-1 & 1 & 0 \end{array}\right)$
	\end{itemize}
		
		We now proceed to analyze the impact of consistency on the admissibility of the loci.
		
		\subsection{Unstability of $\mathcal{R}PC_n$ Under {{}, procedures making consistent a PC matrix}}
		
		Let us examine the effect of a consistency procedure, specifically the orthogonal projection method, on ranking. This method allows us to obtain consistent PC matrices that may or may not preserve the $\mathcal{R}$-condition. Specifically, we examine whether the {{}, chosen procedure making consistent a PC matrix} can move a matrix from $\mathcal{R}PC_n$ to a matrix that no longer satisfies the $\mathcal{R}$-condition. For simplicity, we apply this procedure to a selected consistency method described in \cite{KO1997, KSS2020}.
		
		\begin{Theorem} Let $(a_{i,j})_{(i,j) \in \mathbb{N}_n^2} \in PC_n$ and $(b_{i,j})_{(i,j) \in \mathbb{N}_n^2} \in CPC_n$ be the consistent PC matrix obtained from $(a_{i,j})_{(i,j) \in \mathbb{N}_n^2}$ by the orthogonal projection method. Then:
			
			\begin{itemize} \item The matrix $(a_{i,j})_{(i,j) \in \mathbb{N}_n^2}$ may satisfy the $\mathcal{R}$-condition, while $(b_{i,j})_{(i,j) \in \mathbb{N}_n^2} \in \overline{\mathcal{R}}CPC_n$. \item The matrix $(a_{i,j})_{(i,j) \in \mathbb{N}_n^2}$ may satisfy the $\mathcal{R}$-condition and belong to an admissible locus, while $(b_{i,j})_{(i,j) \in \mathbb{N}_n^2} \in \overline{\mathcal{R}}CPC_n$. \item The matrix $(a_{i,j})_{(i,j) \in \mathbb{N}_n^2}$ may satisfy the $\mathcal{R}$-condition and belong to an admissible locus, while $(b_{i,j})_{(i,j) \in \mathbb{N}_n^2}$ also satisfies the $\mathcal{R}$-condition but belongs to a different admissible locus. \end{itemize} \end{Theorem}
		 \noindent
		\textbf{Proof.}
		We produce all the announced counter-examples in $PC_3.$
		Let us now produce our counter-examples. 
		\begin{itemize}
			\item Let us consider the PC matrix $$\left(\begin{array}{ccc} 1 & e & e^{-1} \\
				e^{-1} & 1 & e \\
				e & e^{-1} & 1 \end{array}\right).$$ It satisfies the $\mathcal{R}-$ condition but it is not in an admissible locus. The matrix obtained is $$\left(\begin{array}{ccc} 1 & 1 & 1   \\
				1 & 1 & 1 \\
				1 & 1 & 1 \end{array}\right),$$ which does not satisfy the $\mathcal{R}-$condition. 
			\item Let us consider the PC matrix $$\left(\begin{array}{ccc} 1 & e & e^{3} \\
				e^{-1} & 1 & e^{-1} \\
				e^{-3} & e & 1 \end{array}\right).$$ It satisfies the $\mathcal{R}-$ condition and it is in an admissible locus. The matrix obtained is $$\left(\begin{array}{ccc} 1 & 1 & 1   \\
				1 & 1 & e^{-2} \\
				1 & e^{2} & 1 \end{array}\right),$$ which does not satisfy the $\mathcal{R}-$condition.
			\item Let us consider the PC matrix $$\left(\begin{array}{ccc} 1 & e & e^{-9} \\
				e^{-1} & 1 & e^{-4} \\
				e^{-9} & e^{4} & 1 \end{array}\right).$$ It satisfies the $\mathcal{R}-$ condition and its characteristic ranking matrix is 
			
			$$\left(\begin{array}{ccc} 0 & 1 & -1 \\
				-1 & 0 & -1 \\
				1 & 1 & 0 \end{array}\right).$$ The matrix obtained is $$\left(\begin{array}{ccc} 1 & e^{-1} & e^{-7}   \\
				e & 1 & e^{-6} \\
				e^7 & e^7 & 1 \end{array}\right),$$ and its characteristic ranking matrix is $$\left(\begin{array}{ccc} 0 & -1 & -1 \\
				1 & 0 & -1 \\
				1 & 1 & 0 \end{array}\right),$$ which shows that we have changed of admissible locus during the {{}, procedure making consistent a PC matrix.}
		\end{itemize}
	\section{On a procedure leading to strict ranking} \label{s3}
	
	We have seen that an existing {{}, procedure making consistent a PC matrix} is not adapted to the conservation of the characteristic ranking matrix. We now propose a method that will always produce a consistent PC matrix satisfying the $\mathcal{R}$-condition, and consequently, a consistent PC matrix that will produce a strict ranking of the items. For this, we have to exclude from the initial set of pairwise comparisons matrices the set $\overline{\mathcal{R}}CPC_n$ of the study. Indeed, on this set, there is no need of {{}, procedures making consistent a PC matrix}, and there are already items that have equal rank. Therefore, we work in the set of \emph{admissible pairwise comparisons matrices} defined by
	$$\mathcal{A}PC_n = PC_n - \overline{\mathcal{R}}CPC_n.$$
	We propose to build a functional 
	$$\Phi:\mathcal{A}PC_n \rightarrow \R_+$$
	such that a PC matrix \( A \) is consistent \emph{and} satisfies the $\mathcal{R}$-condition if and only if
	$$\Phi(A) = 0.$$
	\begin{rem}
		Inconsistency indicators \cite{KMal2017} have the same kind of property: given \( ii \) an inconsistency indicator, the PC matrix \( A \) is consistent if and only if \( ii(A) = 0. \)
	\end{rem}
	
	\begin{Theorem}
		Let \( ii \) be an inconsistency indicator. The map
		$$\Phi : \mathcal{A}PC_n \rightarrow \R$$ defined by 
		$$ \Phi((a_{i,j})_{(i,j)\in \N_n^2}) = \left(\prod_{(i,j)\in \N_n^2, i<j}\frac{ii(A)}{(\log(a_{i,j}))^2 + (ii(A))^{{{},n(n+1)/2}}}\right)\sum_{(i,j)\in \N_n^2, i<j} ((\log(a_{i,j}))^4+1)$$
		is a functional
		\begin{itemize}
		\item {{}, with domain equal to $\mathcal{A}PC_n,$}
			\item which is \( \R_+ \)-valued,
			\item which vanishes only on consistent PC matrices that satisfy the \( \mathcal{R} \)-condition.
		\end{itemize}
	\end{Theorem}

	\noindent
	{{}, \textbf{Proof.}
	Since $ii$ is $\R_+-$ valued, then $\Phi$ is $\R_+-$valued. Let us now prove the main difficulty of the theorem, namely: 
	\begin{itemize}
	\item if $A \in PC_n, \, A \notin \mathcal{A}PC_n,$ then $\Phi(A)$ is not well-defined
	\item if $A \in \mathcal{A}PC_n,$ $\Phi(A)=0 \Leftrightarrow A\in CPC_n.$
	\end{itemize}
	For this, we first remark that $$\sum_{(i,j)\in \N_n^2, i<j} ((\log(a_{i,j}))^4+1) \geq \frac{n(n-1)}{2}.$$ For the first factor,let us analyze one case after another.
	\begin{itemize}
	\item If $A \notin \mathcal{A}PC_n,$, then $A \in \overline{\mathcal{R}}CPC_n,$ this implies that
	\[ \left\{ \begin{array}{l} ii(A)=0 \\
	\exists (i,j)\in \N_n^2, i<j \hbox{ and } \log(a_i,j)=0 \end{array} \right.\]
	therefore at least one of the factors of the product $$ \prod_{(i,j)\in \N_n^2, i<j}\frac{ii(A)}{(\log(a_{i,j}))^2 + (ii(A))^{n(n+1)/2}}$$ is ill defined (roughly speaking, of the type $\frac{0}{0}.$
	Let us now assume that there is only one coefficient $a_{i,j}$ such that $\log a_{i,j} = 0.$ Then, for $0<ii(A)<1,$

\begin{align*}
\prod_{\substack{(i,j)\in \{1,\ldots,n\}^2\\ i<j}}
\frac{\ii(A)}{(\log a_{i,j})^2 + \bigl(\ii(A)\bigr)^{\frac{n(n+1)}{2}}}
&\geq
\left(
\prod_{\substack{(k,l)\in \{1,\ldots,n\}^2\\ k<l,\ (k,l)\neq(i,j)}}
\frac{1}{(\log a_{k,l})^2}
\right)
\left(
\frac{\bigl(\ii(A)\bigr)^{\frac{n(n-1)}{2}}}{\bigl(\ii(A)\bigr)^{\frac{n(n+1)}{2}}}
\right)
\\[0.5em]
&= \frac{C}{\ii(A)},\qquad
\text{where }\\[0.5em]
& C=\prod_{\substack{(k,l)\in \{1,\ldots,n\}^2\\ k<l,\ (k,l)\neq(i,j)}}
\frac{1}{(\log a_{k,l})^2}\ \in \R_{+}^{\ast}.
\end{align*}

\noindent
Therefore, if there is only one term $a_{i,j}$ such that $a_{i,j}=1$, then $\Phi(A)$ is not well defined.

\medskip

The same holds if more coefficients are equal to $1$: by the same reasoning, if $K>1$ is the number
of off-diagonal coefficients $a_{i,j}$ equal to $1$ in $A$, then
\begin{align*}
\prod_{\substack{(i,j)\in \{1,\ldots,n\}^2\\ i<j}}
\frac{\ii(A)}{(\log a_{i,j})^2 + \bigl(\ii(A)\bigr)^{\frac{n(n+1)}{2}}}
&\geq \frac{C'}{\bigl(\ii(A)\bigr)^{\frac{K\,n(n+1) - n(n-1)}{2}}},\qquad
\text{with } C'\in \R_{+}^{\ast}.
\end{align*}

\noindent
This shows that $\Phi(A)$ is not defined whenever $A\notin \mathcal{A}PC_n$.

	\item If $A \in \mathcal{A}PC_n - CPC_n,$ then $\ii(A) > 0$ therefore $\Phi(A)$ is well-defined and $\Phi(A)>0.$
	\item If $A \in \mathcal{A}PC_n olor{red}ap CPC_n,$ then $\forall (i,j) \in \N_n^2,$ $i<j \Rightarrow a_{i,j} \neq 1.$ Therefore, on one side, $\sum_{(i,j)\in \N_n^2, i<j} ((\log(a_{i,j}))^4+1) > \frac{n(n-1)}{2}>0.$ On the other side, each factor of the product $ \prod_{(i,j)\in \N_n^2, i<j}\frac{ii(A)}{(\log(a_{i,j}))^2 + (ii(A))^{n(n+1)/2}}$ is well defined and reads as $\frac{0}{\log(a_{i,j})^2} = 0$ which shows that $\Phi(A)=0.$
	\end{itemize}
	This ends the proof.}
	\vskip 12pt

{{},
\begin{rem}
If $(A_n)$ is a sequence of (non consistent) PC matrices such that every $(1,2)$ coefficient is equal to 1, and such that $\lim \ii(A_n)=0$ then $\lim \Phi(A_n)=+\infty$ by the calculations led in last proof. This seems to indicate that the mapping $\Phi,$ which is obviously continuous and piecewise smooth on $\mathcal{A}PC_n$ if $\ii$ is continuous and piecewise smooth on $PC_n$, is divergent at the neighborhood of $\overline{\mathcal{R}}CPC_n.$ The same way,  on another sequence $(B_n)$ in $\mathcal{A}PC_n,$ if $\ii(B_n)$ is constant and strictly positive, and if only the (1,2)-term diverges to $+ \infty$, then $\Phi(B_n) = O((\log(b_{1,2}))^2)$ therefore $\lim \Phi(B_n) = +\infty.$ These short informal computations enable us to conjecture that most minimization algorithms applied to $\Phi$ will converge to a consistent PC matrix with strict ranking, without generating divergent sequences.    
\end{rem}}
	Within this functional, most classical {{}, procedures making consistent a PC matrix} fail to be generalized to a concrete method leading to its minimization, except the gradient method which may produce an efficient approach, adapting the work \cite{MMC2023} initially developed on inconsistency indicators to the map \( \Phi \).  
	The full study of the functional \( \Phi \), its regularity, and the corresponding gradient method is out of the scope of this paper and is left as an open question.
		\section{Conclusion} 
		
		In this work, we proposed a ranking method on a PC matrix satisfying the \( \mathcal{R}- \) condition without assuming consistency. This change of paradigm, compared to classical procedures, seemingly produce an uncompatible approach with the search for consistency, or at least an uncompatible {{}, approach} with \emph{some} methods, for which we proved that the ranking is violated {{}, by the procedures making a PC matrix consistent.} After these investigations,  we defined a functional \( \Phi \) that, when equal to zero, guarantees both the consistency of the PC matrix and the satisfaction of the \( \mathcal{R}- \) condition.  Finally, the in-depth study of the functional \( \Phi \) and the associated optimization methods remains an open question.
		
		But let us try to understand better what is psychologically underlying the concepts of rankings and consistency. Ranking intends to produce preferences, while the requirement of (at least approximate) consistency is required by the necessary coherence of {{}, assessments.} However, the human mind does not make rankings by numbers, but more by fuzzy (linguistic) values that are difficult to  universally traduce into numbers (see e.g. \cite{MPFMM2021}). Therfore, more than a {{}, consistent ranking} obtained by an arbitrarily chosen procedure, maybe a more complex structure than $\R_+^*$ is required for evaluation in human-related problems, while numbers are sufficient for e.g. wireless technologies. This refinement of viewpoint may split the {{},PC} method, actually so widely spread, to refined, specialized approaches adapted to complex problems, in particular to those with psychological issues. The principal bundle approach with non-abelian Lie groups \cite{Ma2018-3}, and more deeply  with some of their generalizations, potentialy propose such a setting, even if our own opinion is quite pessimistic for such a goal since the human mind is still far from being modelized efficiently (at least to our best knowledge). 
		\section*{Appendix: On finite configuration spaces}
		
		Following \cite{FH2001,M2017}, we set a locally compact manifold $N$, identified with the set of Dirac measures $\left\{ \delta_x \, | \, x \in N \right\}.$
		Let $I$ be a set of indexes that is assumed here to be a finite subset of $\N$ (typically, $I = \N_k = \{1,...,k\}$).
		We define the \textbf{ordered} 
		configuration spaces.
		
		\begin{eqnarray*} O\Gamma^k & = & \lbrace (x_1, ... , x_k) \in N^k 
			\hbox{ such that, if } i \neq j, \quad x_i \neq x_j \rbrace \\
			O\Gamma &=&  \coprod_{k \in \N^*} O\Gamma^k.\end{eqnarray*}
		
		The general configuration spaces are not ordered. For $k \in \N^*,$ let $\mathfrak{S}_k$ be 
		the group of bijections on $\N_k = \{1,\cdots,k\}$. We 
		can define the action: 
		\begin{eqnarray*} \mathfrak{S}_k \times O\Gamma^k & \rightarrow & O\Gamma^k \\
			(\sigma, (x_1, ... , x_k) ) &   \mapsto & (x_{\sigma(1)}, ... , x_{\sigma(k)}). \end{eqnarray*}
		
		Then, we define general configuration spaces:
		$$\Gamma^k = O\Gamma^k  / \mathfrak{S}_k \hbox{ and }
		\Gamma =   \coprod_{k \in \N^*} \Gamma^k . $$
		The manifold $O`\Gamma$ is a $k!-$covering of the manifold $\Gamma$ and moreover: 
		\begin{itemize}
			\item $O`\Gamma^n$ is isomorphic to $\Gamma^n \times \mathfrak{S}_n$ if $N = \R$,
			\item if $n$ is connected and $dim(N) > 1,$ the covering $$\pi_O: O\Gamma^k \rightarrow \Gamma^k$$ is non trivial. 
		\end{itemize}
		{{}, Since} $O\Gamma^k$ is an open subset of $N^k,$ for any $(x_1,\cdots,x_k ) \in O\Gamma^k,$ we have that 
		$$T_{(x_1,\cdots,x_k)}O\Gamma^k = T_{x_1}N \times \cdots \times T_{x_k} N$$ and, by the $\mathfrak{S}_k-$covering of $\Gamma^k$ already defined, we also get that, $\forall x \in \Gamma^k, $
		$$T_x \Gamma^k =  T_{x_1}N \times \cdots \times T_{x_k} N$$ where $\pi_O(x_1,...,x_k) = x.$
		Let us now concentrate our efforts on $N = \R^n,$ equipped with its natural (constant) Riemannian metric (or Euclidian scalar product) $g$ with Euclidian norm $||.||.$  
		\begin{Definition}
			We define the following metric on $O\Gamma^k :$
			$$g_{(x_1,\cdots,x_k)}( (v_1,\cdots,v_k), (v'_1,\cdots,v'_k)) = \left(\sum_{(i,j)\in  \N_k^2}\frac{1}{ ||x_i - x_j||^2}\right)^2 \sum_{l \in \N_k} g(v_l,v'_l).$$
		\end{Definition}
		
		This metric has the interesting property to be $\mathfrak{S}_k-$invariant, and hence to be also a Riemannian metric on $\Gamma.$ Moreover, 
		
		\begin{Proposition}
			Let {{}, $\gamma \in C^\infty(\R,N^k)$ such that $$\forall t < 1, \gamma(t) \in O\Gamma^k \hbox{ and } \gamma(1) \notin O\Gamma^k.$$}
			Denoting $L\gamma(t)$ the length o$\gamma$ on $[0,t]$ for $t<1$ and for the Riemannian metric $g_O,$ we have $$\lim_{t \rightarrow 1^-} L\gamma(t) = + \infty.$$
		\end{Proposition}
		
		\noindent
		\textbf{Proof.}
		Since $\gamma$ is smooth on {{}, $\R,$} we can assume that $||\dot \gamma|| = 1$ and we have that $$ \int_0^1 \sqrt{g_O(\dot\gamma(t),\dot \gamma(t) )} dt$$ is a divergent improper integral.
		\vskip 12pt
		
		\begin{rem}
			This Riemannian metric on $O\Gamma^k$ is directly inspired from the hyperbolic metric on the Poincar\'e disk \cite{Gro}. To our best knowledge, and to our great surprise, this metric has never been defined and studied, neither in a published paper nor in any preprint available online. We suspect an interesting hyperbolic geometry related to this Riemannian metric, but we leave this differential geometric study as an open question.  
		\end{rem}
		
		\paragraph{\bf Acknowledgements:} J.-P.M. thanks the France 2030 framework programme Centre Henri Lebesgue ANR-11-LABX-0020-01 
		for creating an attractive mathematical environment.


\begin{thebibliography}{150}
			\bibitem{BR2022} Bartl, D.; Rama­k, J.; A new algorithm for computing priority vector of
			pairwise comparisons matrix with fuzzy elements. \emph{ Information Sciences},
			{\bf 615} 103-117 (2022).
			\bibitem{B1938} Benford, F.; The law of anomalous numbers. \emph{Proc. Am. Philos. Soc.} {\bf 78} no 4, 551-572 (1938)
			\bibitem{BR2008} Boz\'oki, S.;  Rapcs\'ak, T.;  On
			Saaty's and
			Koczkodaj's inconsistencies
			of
			pairwise
			comparison
			matrices;
			\textit{J.
				Glob. Optim.}
			\textbf{42} no.2
			(2008)
			157-175 
			\bibitem{CW2004} Choo, E. U.; Wendley, W.C.; A common framework for deriving preferences values from pairwise comparisons matrices. \emph{Comp. Oper. Research} {\bf 256} no1, 783--803 (2004) 
			\bibitem{6} Colomer, J-M.; Ramon Llull: from Ars electionis to social choice theory.
			\emph{Social Choice and Welfare} {\bf 40} no2, 317-328 (2011).
			\bibitem{CW1985} Crawford, R.; Williams, C.; A note on the analysis of subjective judgement matrices. \emph{Journal of Mathematical Psychology}, {\bf 29} 387--405 (1985)
			\bibitem{10} A. Darko, A.;  Chan,  A. P. C.; Ameyaw, E. E.; Owusu,  E. K.;  P\"arn, E.; D. J.
			Edwards,  D. J.; Review of application of analytic hierarchy process (AHP) in construction. \emph{International Journal of Construction Management} {\bf 19} no5, 436-452 (2019)
			\bibitem{EML2024} Ellingsen, A.; Lundholm, D.; Magnot, J-P.;  ``The six blind men and the elephant'': an interdisciplinary selection of measurement features. in:
			Kielanowski, Piotr (ed.) et al., Geometric methods in physics XL, workshop, Bialowieza, Poland, June 20?25, 2023. Cham: Birkhauser. \emph{Trends Math.}, 275-307 (2024). 
			\bibitem{FH2001} Fadell, E.R.; Husseini, S.Y.; \textit{Geometry and topology of configuration spaces} 
			Springer, Berlin (2001)	
			\bibitem{FV2012} Farinelli S., Vasquez S. Gauge invariance, geometry and arbitrage. \emph{J. Invest. Strategies}. {\bf 1} no 2, 2366 (2012).
			\bibitem{F2008} F\"ul\"op, J.; A method for approximating pairwise comparisons matrices by consistent matrices \textit{J. Global Optimization} \textbf{42}, 423-442 (2008)
			\bibitem{14} Gacula Jr., M.C.;  Singh, J.; \emph{Statistical Methods in Food and Consumer Research}. Academic Press (1984).
			\bibitem{18} Hyde, R.A.; Davis, K.; Military applications of the analytic hierarchy
			process. \emph{International Journal of Multicriteria Decision Makin}g, {\bf 2} no3), 267
			(2012)
			\bibitem{Gro} 
			Gromov, M.; \emph{ Metric structures in Riemannian and non-Riemannian spaces}, 2nd ed. Birkauser, (1997).
			\bibitem{I2000} Illinski, K.; Gauge geometry of financial markets. \emph{J. Phys. A, Math. Gen.} {\bf 33} 5-14 (2000).
			\bibitem{KN} Kobayashi, S.; Nomizu, K.; {`\it Foundations of Differential Geometry I and II} Wiley classics library (1963-1969) 	
			\bibitem{K1993} Koczkodaj, W.W.; A new definition
			of consistency
			of pairwise
			comparisons,\textit{Math. Comput.
				Modelling} {\bf 8}
			(1993) 79-84.
			\bibitem{KMal2017}	 Koczkodaj, W.W.;  Magnot, J-P.;  Mazurek, J.; Peters, J.F.;
			Rakhshani, H.; Soltys, M.;  Strza lka, D.; Szybowski, J.;   Tozzi, A.;On normalization of inconsistency indicators
			in pairwise comparisons \textit{Int. J. Approx. Reasoning} \textbf{86} (2017) 73-79.
			\bibitem{KO1997} Koczkodaj, W.; Orlowski, M.; An orthogonal basis for computing a
			consistent approximation to a pairwise comparisons matrix. \emph{Computers
				and Mathematics with Applications} {\bf 34} no10, 41-47 (1997)
			\bibitem{KSS2020} Koczkodaj, W.W.; Smarzewski, R.; Szybowski, J.; On Orthogonal Projections on the Space of Consistent Pairwise Comparisons Matrices \emph{Fundamenta Informaticae}, {\bf 172}, no. 4  379-397 (2020)
			\bibitem{KGDW2015} Kulakowski, K. Grobler-Dobska, K; Was, J.; Heuristic rating estima-
			tion: geometric approach. \emph{Journal of Global Optimization}, {\bf 62} no3, 529-543
			(2015).
			\bibitem{27}  Kulakowski, K., Mazurek, J., Strada, M.; On the similarity between
			ranking vectors in the pairwise comparison method. \emph{Journal of the Operational Research Society}, {\bf 0} 1-10 (2021).
			\bibitem{LLA2012} Lahby, M.; Leghris, C.; Adib, A.; Network Selection Decision based on handover history
			in Heterogeneous Wireless Networks; \textit{International 
				Journal of 
				Computer 
				Science and 
				Telecommunications} {\bf 3} no 2,  (2012) 21-25
			\bibitem{32} Liberatore, M.J.; Nydick, R. L.; Group decision making in higher education using the analytic hierarchy process. \emph{ Research in Higher Education},
			{\bf 38} no5, 593?614 (1997).
			\bibitem{M2017} Magnot, J-P.; From Configurations to Branched Configurations and Beyond. \emph{Research and Reports in Mathematics} {\bf 1}
			no 1, art. ID 1000105 (2017).
			\bibitem{Ma2018-1} Magnot, J-P.; A mathematical bridge between discretized gauge theories in quantum physics and approximate reasoning in pairwise comparisons \textit{Adv. Math. Phys.} \textbf{2018}, Article ID 7496762, 5 pages (2018).
			\bibitem{Ma2018-3} Magnot, J-P.Â ; On mathematical structures on pairwise comparisons matrices with coefficients in a group arising from quantum gravity {\it Helyion} {\bf 5} e01821 (2019) 
			\bibitem{MMC2023} Magnot, J-P.; Mazurek, J.; Cernanova, V.; A gradient method for inconsistency reduction of pairwise comparisons matrices \emph{Int. J. Approx. Reasonning} {\bf 152} 46-58 (2023) 
			\bibitem{M2024} Magnot, J-P.; On random pairwise comparisons and their geometry. \emph{J. Appl. Anal.} In press. 
			\bibitem{MPFMM2021} Mazurek, J.; Perez Rico, C.; Fernandez, C.; Magnot, J-P.; Magnot, T.; The 5-Item Likert Scale and Percentage Scale Correspondence with Implications for the Use of Models with (Fuzzy) Linguistic Variables. Journal of Quantitative Methods for Economics and Business Administration Volume 31 3-16 (June 2021)
			\bibitem{37} Peterson, G.L.; Brown, T.C.; Economic valuation by the method of paired comparison, with emphasis on evaluation of the transitivity axiom.
			\emph{Land Economics} 240-261 (1998)
			\bibitem{S1977} Saaty, T.; A scaling methods for priorities in hierarchical structures; \textit{J. Math. Psychol.} \textbf{15} (1977) 234-281
			\bibitem{41} Sasaki, Y.; Strategic manipulation in group decisions with pairwise comparisons: A game theoretical perspective. \emph{ European Journal of Operational
				Research} {\bf 304} no3, 1133-1139 (2023).
			\bibitem{42} Taylor, A.D.; \emph{Social Choice and the Mathematics of Manipulation}. Outlooks. Cambridge University Press (2005).
			\bibitem{43} Thurstone, L.L.; The Method of Paired Comparisons for Social Values. \emph{Journal of Abnormal and Social Psychology}, pages 384-400 (1927).
			\bibitem{45}  Urbaniec, M.; Soltysik, M.; Prusak, A.; Kurakowski, K.; 
			Wojnarowska, M.; Fostering sustainable entrepreneurship by busi-
			ness strategies: An explorative approach in the bioeconomy. {Business Strategy and the Environment} {\bf 31} no1, 251-267 (2022).
			\bibitem{Wh} 
			Whitney, H., Geometric Integration Theory, Princeton University Press, Princeton, NJ, 1957.
		\end{thebibliography}
\end{document}